# Thermal entanglement in a two-qutrit system with nonlinear coupling under nonuniform external magnetic field

Iman Sargolzahi, Sayyed Yahya Mirafzali, Mohsen Sarbishaei
Department of Physics, Ferdowsi University of Mashhad, Iran

**Abstract**

We study the thermal entanglement of a 2-qutrit spin chain with nonlinear coupling in the presence of nonuniform magnetic field. Thermal entanglement of an arbitrary (finite dimensional) *m*-partite system vanishes at some finite threshold temperature $T_s$. We investigate the dependence of $T_s$ on the system's parameters, i.e. the nonlinear coupling and the magnetic field, for this 2-qutrit system. In addition, we compare two lower bounds of I-concurrence for this system and also study its dense coding capacity as a function of system's parameters.

## 1. Introduction

A bipartite system is called entangled if its density operator can not be written as
$$\rho_{AB} = \sum_i p_i \rho_i^A \otimes \rho_i^B \qquad (1)$$
where $\rho_i^A = |\psi_i\rangle_A \langle\psi_i|$, $\rho_i^B = |\phi_i\rangle_B \langle\phi_i|$, $p_i \geq 0$, $\sum_i p_i = 1$ and $|\psi_i\rangle_A$, $|\phi_i\rangle_B$ are normalized pure states of subsystems A and B, respectively. The generalization of the above definition to *m*-partite ($m \geq 2$) systems is straightforward. An *m*-partite system is separable (entangled) if its density operator can (can not) be written as $\sum_i p_i \rho_i^{(1)} \otimes \cdots \otimes \rho_i^{(m)}$, where $p_i \geq 0$, $\sum_i p_i = 1$ and $\rho_i^{(j)}$ are density operators of the subsystem j.

Entanglement plays a key role in quantum information theory, in applications such as teleportation, dense coding and so on [1, 2]. In quantum information protocols usually qubits are used; however entangled qutrits are also of interest [3- 9]. For example Durt *et al.* [6] have introduced a cryptography protocol with entangled qutrits which is more robust than its entangled qubits' counterpart.

Entanglement is usually a fragile property. So creating stable entanglement is desirable. A system at thermal equilibrium can produce stable entanglement. Such a system is described by:
$$\rho(T) = \exp(-\beta H)/Z \qquad (2)$$
where *H* is the Hamiltonian, $Z = Tr[\exp(-\beta H)]$ is the partition function, $\beta = 1/(k_B T)$, *T* is the temperature and $k_B$ is the Boltzmann's constant which we set $k_B = 1$. If some of the Hamiltonian's eigenvectors are entangled states, then entanglement may occur naturally for some intervals of $T$. Such a kind of entanglement is called thermal entanglement [10].

One can define $T_s$ as the temperature that $\rho(T)$ is *separable* for all $T \geq T_s$ [11]. It can be shown that $T_s$ is finite (see Sec.4). In this paper we will study the thermal entanglement of a 2-qutrit spin chain by using the negativity as the measure of entanglement. The interesting feature of this system, as we will see in Sec. 7, is that increasing the absolute value of the nonlinear coupling can increase both $T_s$ and the entanglement of the ground state.

Dense coding capacity of a bipartite state is a measure of the amount of information that can be stored in it by manipulating only one part of the total system (see Sec. 3). As we will see in Sec. 9, the effect of increasing the absolute value of the nonlinear coupling on the usefulness of the system for dense coding is similar to its effect on the negativity of the system.

## 2. Concurrence and Negativity

*Negativity* is a measure of entanglement for bipartite systems and is defined as [12]:

$$N(\rho) = \frac{\|\rho^{T_j}\|_1 - 1}{2} \tag{3}$$

where $\rho^{T_j}$ denotes the partial transpose of $\rho$ with respect to the *j*th subsystem ($j$=1,2) and $\|A\|_p$, for any operator $A$, is defined as $\left(\sum_i |a_i|^p\right)^{\frac{1}{p}}$ where $|a_i|$ are singular values of $A$, i.e. $|a_i|^2$ are eigenvalues of $A^\dagger A$. It can be shown that $N(\rho)$ is the absolute value of the sum of the negative eigenvalues of $\rho^{T_j}$. Negativity can detect all entangled states for any $2 \times 2$ and $2 \times 3$ dimensional bipartite system [13], but it can not reveal all entangled states for higher-dimensional systems.

For a bipartite $2 \times 2$ dimensional system, *concurrence* is defined as [14]:

$$C(\rho) = \max\left\{\Lambda_1 - \sum_{j>1}^{4} \Lambda_j, 0\right\} \tag{4}$$

where $\Lambda_j$ are the square roots of the eigenvalues of the matrix $R = \rho(\sigma_y \otimes \sigma_y)\rho^*(\sigma_y \otimes \sigma_y)$ in decreasing order, where $\sigma_y$ is the second Pauli matrix, $\rho^*$ is the complex conjugate of the matrix $\rho$ and the standard basis, i.e. $\{|00\rangle, |01\rangle, |10\rangle, |11\rangle\}$, is used to represent the $\rho$. It can be shown that the entanglement of formation of $\rho$ is a monotonically increasing function of $C(\rho)$; so $C(\rho)$ can be considered as an independent entanglement measure [14].

A generalization of concurrence to higher dimensional systems is called *I-concurrence*. For a pure bipartite state $|\psi\rangle \in H_A \otimes H_B$, where $H_A$ ($H_B$) is the Hilbert space of the subsystem $A$ ($B$), I-concurrence of $|\psi\rangle$ is defined as [15]:

$$C(|\psi\rangle) = \sqrt{2[1 - Tr(\rho_A^2)]} \qquad (5)$$

where $\rho_A$ is the reduced density operator of the subsystem $A$, i.e. $\rho_A \equiv Tr_B(|\psi\rangle\langle\psi|)$. $Tr(\rho_A^2) = 1$ if and only if $\rho_A$ is pure; i.e. iff $|\psi\rangle$ is a product state. So $C(|\psi\rangle) = 0$ if and only if $|\psi\rangle$ is separable. The extension of the above definition of I-concurrence to mixed states is straightforward. For a mixed state $\rho : H_A \otimes H_B \to H_A \otimes H_B$, $C(\rho)$ is defined as [15]:

$$C(\rho) = \min_{\{p_i, |\psi_i\rangle\}} \sum_i p_i C(|\psi_i\rangle) \qquad (6)$$

where the minimum is taken over all decompositions of $\rho$ into pure states: $\rho = \sum_i p_i |\psi_i\rangle\langle\psi_i|$ with $p_i \geq 0$ and $\sum_i p_i = 1$. From Eqs. (5) and (6) it can be seen that $\rho$ is separable if and only if $C(\rho) = 0$; as $C(\rho) = 0$ iff there is a decomposition of $\rho$ into pure product states. Unfortunately $C(\rho)$ can not be computed in general. However some lower bounds are introduced for it. Chen *et al.* [16] have shown that

$$C(\rho) \geq \sqrt{\frac{8}{m(m-1)}} N(\rho) \qquad (7)$$

where $m \equiv \min(d_A, d_B)$, $d_A$ is $\dim(H_A)$ and $d_B$ is $\dim(H_B)$. Mintert *et al.* [15] have introduced another lower bound for $C(\rho)$:

$$C(\rho) \geq \max\left\{\zeta_1 - \sum_{j>1} \zeta_j, 0\right\} \qquad (8)$$

where $\zeta_j$ are singular values of a matrix $\tau$ in decreasing order. $\tau$ is defined as

$$\tau = \sum Z_\alpha T^\alpha \qquad (9)$$

where $Z_\alpha$ are arbitrary complex numbers which satisfy the constraint $\sum_\alpha |Z_\alpha|^2 = 1$ and matrices $T^\alpha$ are constructed as below.

Assume that $\{|0_A\rangle, \ldots, |(d_A - 1)_A\rangle\}$ and $\{|0_B\rangle, \ldots, |(d_B - 1)_B\rangle\}$ are orthonormal bases for $H_A$ and $H_B$ respectively. Now we define $|t_{m_A}\rangle \equiv |j_A\rangle|k_A\rangle - |k_A\rangle|j_A\rangle$ (with $j = 0, \ldots, d_A - 2$, $k = j+1, \ldots, d_A - 1$ and so $m_A = 1, \ldots, d_A(d_A - 1)/2$) as an orthogonal basis for the antisymmetric subspace of $H_A \otimes H_A$. Similarly, we can construct $\{|t_{m_B}\rangle\}$.

We now define $|\chi_\alpha\rangle \in (H_A \otimes H_A) \otimes (H_B \otimes H_B) = (H_A \otimes H_B) \otimes (H_A \otimes H_B)$ as $|\chi_\alpha\rangle = |t_{m_A}\rangle \otimes |t_{m_B}\rangle$, $\alpha = 1, \ldots, n \equiv \frac{d_A(d_A - 1)d_B(d_B - 1)}{4}$. Now we can construct the matrices $T^\alpha$ with the elements:

$$T^\alpha_{jk} = \sqrt{\lambda_j \lambda_k} \langle \chi_\alpha | \Phi_j \rangle | \Phi_k \rangle \qquad (10)$$

where $|\Phi_i\rangle$ are the eigenstates of $\rho$ and $\rho|\Phi_i\rangle = \lambda_i |\Phi_i\rangle$.

In this paper, we will use the negativity as the measure of entanglement for our 2-qutrit system. However, since it can not reveal all entangled states in a 2-qutrit system, it seems interesting to compare it with other entanglement measures in this system. We will compare the negativity with a special class of Mintert *et al.*'s lower bounds of I-concurrence in Sec. 8.

## 3. Dense Coding Capacity

Suppose that Alice (A) wants to send to Bob (B) one of the four classical information $\{0 = 00, 1 = 01, 2 = 10, 3 = 11\}$. Classically, Alice can do so by sending 2 bits to Bob; but, interestingly, sending only a qubit is sufficient if they have initially share a singlet state:

$$|\psi^-\rangle = \frac{1}{\sqrt{2}}\left(|0_A 1_B\rangle - |1_A 0_B\rangle\right) \qquad (11)$$

This interesting application of entangled states is called *dense coding* and the procedure is as follows [1, 2, 16]. According to the choice of the classical information which Alice wants to send to Bob, she performs one of the four unitary transformations $\{U_{0,0} = I, U_{0,1} = \sigma_x, U_{1,0} = \sigma_z, U_{1,1} = -i\sigma_y\}$ on her qubit, where $I$ is the identity operator and $\sigma_{x/y/z}$ are the Pauli operators. So the singlet state transforms to one of the following states:

$$\begin{aligned} 00: I \otimes I |\psi^-\rangle &= |\psi^-\rangle \quad, \quad 01: \sigma_x \otimes I |\psi^-\rangle = -|\varphi^-\rangle \\ 10: \sigma_z \otimes I |\psi^-\rangle &= |\psi^+\rangle \quad, \quad 11: -i\sigma_y \otimes I |\psi^-\rangle = |\varphi^+\rangle \end{aligned} \qquad (12)$$

where $|\psi^\pm\rangle \equiv \frac{1}{\sqrt{2}}(|01\rangle \pm |10\rangle)$ and $|\varphi^\pm\rangle \equiv \frac{1}{\sqrt{2}}(|00\rangle \pm |11\rangle)$ are the four *Bell states*. Then Alice sends her qubit to Bob, and since the Bell states are orthonormal, he can distinguish between them by measuring the 2-qubit system in the Bell basis and find out the classical information sent by Alice.

Dense coding protocol is, in fact, an example of encoding the classical information in a quantum system and then decoding this information by measuring it. In the dense coding protocol, the encoding procedure is restricted to manipulating only a qubit of the total 2-qubit system; so the amount of information that can be stored and then extracted from this 2-qubit system is related to the amount of initial entanglement between the two qubits. The above argument leads us to the definition of the dense coding capacity of an entangled system and, as we will see, this capacity is related to a measure of entanglement: the *relative entropy of entanglement*.

Let us first generalize the standard dense coding protocol to include an arbitrary entangled state instead of the singlet state and also to include an arbitrary set of $\{U_{x,y}\}$ instead of $\{U_{0,0} = I, U_{0,1} = \sigma_x, U_{1,0} = \sigma_z, U_{1,1} = -i\sigma_y\}$ [16]. Suppose that Alice wants to send to Bob one of the $d^2$ classical information $(x,y)$, where $x=0,\ldots,(d-1)$ and $y=0,\ldots,(d-1)$, while they have initially shared an arbitrary 2-qudit entangled state $\rho_{AB}$. In order to send the classical information $(x,y)$, Alice performs a unitary transformation $U_{x,y}$ on her qudit. So the initially shared state $\rho_{AB}$ transforms to

$$\rho_{x,y} = U_{x,y} \otimes I \rho_{AB} U^\dagger_{x,y} \otimes I \tag{13}$$

Then Alice sends her qudit to Bob who tries to find out the classical information $(x,y)$ by performing a measurement on the total 2-qudit system. Now we can define the dense coding capacity of $\rho_{AB}$ by means of the Holevo bound [1]. Assume that Alice produces the state $\rho_{x,y}$ with the probability $p_{x,y}$. Now Bob has the state $\bar{\rho} = \sum_{x,y} p_{x,y} \rho_{x,y}$ in hand and tries to find out the real ensemble produced by Alice, i.e. $\{p_{x,y}, \rho_{x,y}\}$, by performing a measurement on $\bar{\rho}$. The average classical information that Bob can gain, or in other words the average classical information that can be transmitted by sending a qudit from Alice to Bob, is bounded from above by the *Holevo $\chi$ quantity* [1]:

$$\chi(\{p_{x,y}, \rho_{x,y}\}) = S(\bar{\rho}) - \sum_{x,y} p_{x,y} S(\rho_{x,y}) \tag{14}$$
$$= S(\bar{\rho}) - S(\rho_{AB})$$

where $S(\rho) \equiv -Tr(\rho \log_2 \rho)$ and we have used the fact that $S(\rho_{x,y}) = S(\rho_{AB})$, as $\rho_{x,y}$ is a unitary transformation of $\rho_{AB}$. The *dense coding capacity* of $\rho_{AB}$ is defined as [18]:

$$C_{DC}(\rho_{AB}) = \max_{p_{x,y}, \rho_{x,y}} \chi(\{p_{x,y}, \rho_{x,y}\}) \tag{15}$$

It can be shown [19] that the maximum of $\chi(\{p_{x,y}, \rho_{x,y}\})$ is achieved when $p_{x,y} = \frac{1}{d^2}$ for all $(x,y)$ and

$$U_{x,y} = \sum_{j=0}^{d-1} e^{i\frac{2\pi}{d}jx} |j+y \bmod(d)\rangle\langle j| \tag{16}$$

where $\{|j\rangle\}$ is an orthonormal basis for the subsystem A. In addition, under the above circumstances we have [19]:

$$\bar{\rho} = \sum_{x,y} p_{x,y} \rho_{x,y} \equiv \bar{\rho}^* = \frac{1}{d} I_A \otimes \rho_B \tag{17}$$

where $I_A$ is the identity operator of the subsystem A and $\rho_B \equiv Tr_A(\rho_{AB})$ is the reduced density operator of the subsystem B. So:

$$C_{DC}(\rho_{AB}) = S(\overline{\rho}^*) - S(\rho_{AB})$$
$$= -Tr_{AB}\{[\frac{I_A}{d} \otimes \rho_B][I_A \otimes \log_2(\frac{\rho_B}{d})]\} - S(\rho_{AB}) \quad (18)$$
$$= -Tr_B\{\rho_B[\log_2(\rho_B) - \log_2 d]\} - S(\rho_{AB})$$
$$= \log_2 d + S(\rho_B) - S(\rho_{AB})$$

For a pure product state $\rho_{AB;P} = |\psi\rangle_P\langle\psi|$ with $|\psi\rangle_P = |\psi_A\rangle \otimes |\psi_B\rangle$ we have $S(\rho_{AB;P}) = S(\rho_B) = 0$; so $C_{DC}(\rho_{AB;P}) = \log_2 d$ which is the same as the classical case; i.e. if we have a 2-dit system, manipulating only one of the two dits can store, at most, $\log_2 d$ bits of information, or equivalently one dit of information, on the 2-dit system. But for the maximally entangled state $\rho_{AB;ME} = |\psi\rangle_{ME}\langle\psi|$ with $|\psi\rangle_{ME} = \frac{1}{\sqrt{d}}\sum_{i=0}^{d-1}|i_A\rangle|i_B\rangle$, where $\{|i_A\rangle\}$ and $\{|i_B\rangle\}$ are orthonormal bases for the subsystems A and B respectively, we have: $S(\rho_{AB;ME}) = 0$, $S(\rho_B) = \log_2 d$ and so $C_{DC}(\rho_{AB;ME}) = 2\log_2 d = \log_2 d^2$. For the singlet state $C_{DC}(|\psi^-\rangle\langle\psi^-|) = 2$, which implies that we can transmit 2 bits of information by sending only a qubit from Alice to Bob. Also note that for the 2-qubit case, $\{U_{x,y}\}$ in Eq. (16) are the same as those in Eq. (12).

Equation (18) implies that only those states for which $S(\rho_B) - S(\rho_{AB}) > 0$ are useful for dense coding. So we expect that $S(\rho_B) - S(\rho_{AB})$ relates to the amount of entanglement of $\rho_{AB}$. In fact it can be shown that [20]:
$$E_R(\rho_{AB}) \geq \max\{S(\rho_A) - S(\rho_{AB}); S(\rho_B) - S(\rho_{AB})\} \quad (19)$$
where:
$$E_R(\rho_{AB}) \equiv \min_{\sigma_{AB} \in D} S(\rho_{AB} \| \sigma_{AB}) \quad (20)$$
is the *relative entropy of entanglement* [2, 21]. In Eq. (20), $S(\rho\|\sigma) \equiv Tr[\rho(\log_2\rho - \log_2\sigma)]$ is the *relative entropy* of $\rho$ to $\sigma$ [1] and $D$ is either the set of separable states or the set of states which are positive under partial transposition or the set of non-distillable states [18, 19, 2]. So Eq. (19) implies that if either $S(\rho_A) - S(\rho_{AB}) > 0$ or $S(\rho_B) - S(\rho_{AB}) > 0$ (so $\rho_{AB}$ is useful for dense coding) then $\rho_{AB}$ is an entangled NPT distillable state, where NPT states are those states for which $N(\rho) > 0$.

In Sec. 9 we will study the behavior of $S(\rho_{A/B}) - S(\rho_{AB})$ for our 2-qutrit system, as a function of system's parameters.

## 4. Thermal Entanglement Vanishes at a Finite Temperature

Using the fact that the maximally mixed state $\rho = I/d$, where $d$ is the dimension of the Hilbert space of the system and $I$ is the identity operator, is surrounded by separable states [22-25], Fine *et al.* [26] have proved that thermal entanglement in an arbitrary (finite dimensional) bipartite system vanishes at a finite temperature. In fact it is true for an arbitrary (finite dimensional) *m*-partite ($m \geq 2$) system; as has been stated by some authors [11, 27-29]. In this section, we will give a proof of this statement using the result of Ref. [25].

Consider an arbitrary finite dimensional *m*-partite density operator $\rho$: $H_1 \otimes \ldots \otimes H_m \to H_1 \otimes \cdots \otimes H_m$ where $H_i$ ($1 \leq i \leq m$) is the Hilbert space of the i'th subsystem. Gurvits and Barnum [25] have shown that $\rho$ is separable if

$$\left\| \rho - I/d \right\|_2 \leq \left[ 2^{\frac{m-2}{2}} \sqrt{d(d - 2^{-(m-2)})} \right]^{-1} \tag{21}$$

where $d = \dim(H_1 \otimes \cdots \otimes H_m) \equiv d_1 \times \cdots \times d_m$. Since $\left\| \rho - I/d \right\|_2^2 = \left\| \rho \right\|_2^2 - \frac{1}{d}$, we can rewrite Eq. (21) in terms of the *purity* of $\rho$: If

$$P(\rho) \equiv Tr[\rho^2] = \left\| \rho \right\|_2^2 \leq (d - 2^{2-m})^{-1} \tag{22}$$

then $\rho$ is separable.

For a thermal state $\rho$:

$$\rho = \sum_i \lambda_i |\Phi_i\rangle\langle\Phi_i| \quad \text{with} \quad \lambda_i = e^{-\beta E_i}/Z \tag{23}$$

where $H|\Phi_i\rangle = E_i|\Phi_i\rangle$, we have $\dfrac{dP(\rho)}{d\beta} = \sum_{ij} 2\lambda_i^2 \lambda_j (E_j - E_i)$. Since

$$\sum_{ij} \lambda_i \lambda_j E_j (\lambda_i - \lambda_j) \geq 0 \tag{24}$$

then

$$\sum_{ij} \lambda_i^2 \lambda_j E_j \geq \sum_{ij} \lambda_j^2 \lambda_i E_j = \sum_{ij} \lambda_i^2 \lambda_j E_i \quad \Rightarrow \quad \frac{dP}{d\beta} \geq 0$$

Equation (24) is satisfied for all $T$ since when $E_i > E_j$ then $\lambda_i < \lambda_j$. In fact, the equality sign in Eq. (24) holds only when $T = 0$ or $T \to \infty$ or when all $E_j$ are equal (when all $E_j$ are equal then $\rho(T) = I/d$ for all $T$). Neglecting this last unimportant case, $P[\rho(T)]$ is a monotonically increasing function of $\beta$ for $0 < \beta < \infty$. So there is a unique finite temperature $T_*$ for which we have

$$P(T_*) = Tr[\rho^2(T = T_*)] = (d - 2^{2-m})^{-1} \tag{25}$$

Therefore $\rho(T)$ is separable for all $T \geq T_*$. In other words, since $T_*$ is an upper bound of $T_s$, $T_s$ is finite for any (finite dimensional) *m*-partite system.

Now for bipartite systems, Eqs. (21) and (22) reduce to

$$\left\| \rho - I/d \right\|_2 \leq \left[ d(d-1) \right]^{\frac{-1}{2}} \equiv r_{\max} \tag{26-a}$$

and

$$P(\rho) \equiv Tr[\rho^2] \leq (d-1)^{-1} \tag{26-b}$$

respectively. It has been shown that $r_{max}$ is the largest possible radius, in the Hilbert-Schmidt norm (i.e. $\|A\|_2$ for arbitrary operator $A$), around $\frac{I}{d}$ which includes only separable states [24]. So one may conjecture that at least for bipartite systems $T_*$ is a good upper bound for $T_s$, but it is not so, at least, for systems which have been studied in Ref. [11]. Note that $T_*$ is only a function of $\rho$'s eigenvalues and is independent of its eigenvectors, while entanglement is related to both eigenvalues and eigenvectors of $\rho$. This may be the reason why $T_*$ is a poor upper bound of $T_s$ (at least for systems which have been studied in Ref. [11] ).

After knowing that $T_s$ is finite, the other interesting question is how it depends on the system's parameters. We will deal with this question for our 2-qutrit system in Sec. 7.

## 5. The Model Hamiltonian

We will study the thermal entanglement of a bipartite spin-1 chain. In the presence of nonuniform external magnetic field (along the z-axis) the Hamiltonian of this system reads:
$$H = J(\vec{S}_1 \cdot \vec{S}_2) + K(\vec{S}_1 \cdot \vec{S}_2)^2 + B_1 S_{1z} + B_2 S_{2z} \qquad (27)$$
where $J$ is the linear coupling, $K$ is the nonlinear coupling, $B_i$ is the magnetic field at site i and $\vec{S}_i \equiv (S_{ix}, S_{iy}, S_{iz})$ which

$$S_{ix} = \tfrac{1}{\sqrt{2}}\begin{pmatrix} 0 & 1 & 0 \\ 1 & 0 & 1 \\ 0 & 1 & 0 \end{pmatrix}, \quad S_{iy} = \tfrac{1}{\sqrt{2}}\begin{pmatrix} 0 & -i & 0 \\ i & 0 & -i \\ 0 & i & 0 \end{pmatrix}, \quad S_{iz} = \begin{pmatrix} 1 & 0 & 0 \\ 0 & 0 & 0 \\ 0 & 0 & -1 \end{pmatrix}$$

The Hamiltonian in Eq. (27) can be implemented physically by using an optical lattice setup consisting of two wells with one spin-1 atom in each well in the presence of external electric and magnetic fields [30]. External magnetic field produces the term $B_1 S_{1z} + B_2 S_{2z}$ in Eq. (27); but external electric field influences the values of $J$ and $K$ and helps us achieve different values of $J$ and $K$ [30].

The eigenvectors of the Hamiltonian in Eq. (27), in the basis of the eigenstates of $S_{1z} \otimes S_{2z}$ ($S_{1z} \otimes S_{2z} |ij\rangle = i \times j |ij\rangle$), are:

$$\begin{aligned}
|\Phi_1\rangle &= |11\rangle \\
|\Phi_2\rangle &= (a|10\rangle + b|01\rangle)/\sqrt{a^2+b^2} \\
|\Phi_3\rangle &= (b|10\rangle - a|01\rangle)/\sqrt{a^2+b^2} \\
|\Phi_4\rangle &= c_1|00\rangle + d_1|1,-1\rangle + e_1|-1,1\rangle \\
|\Phi_5\rangle &= c_2|00\rangle + d_2|1,-1\rangle + e_2|-1,1\rangle
\end{aligned} \qquad (28)$$

$$|\Phi_6\rangle = c_3|00\rangle + d_3|1,-1\rangle + e_3|-1,1\rangle$$
$$|\Phi_7\rangle = (f|-1,0\rangle + g|0,-1\rangle)/\sqrt{f^2+g^2}$$
$$|\Phi_8\rangle = (g|-1,0\rangle - f|0,-1\rangle)/\sqrt{f^2+g^2}$$
$$|\Phi_9\rangle = |-1,-1\rangle$$

where $a = B_1 - B_2 + \sqrt{(B_1-B_2)^2 + 4J^2}$, b = g = 2J, $f = B_2 - B_1 + \sqrt{(B_1-B_2)^2 + 4J^2}$ and, in general, $\{e_i, d_i c_i, i=1,2,3\}$ should be computed numerically for each $J$, $K$, $B_1$ and $B_2$. The eigenvalues of our Hamiltonian are:

$$E_1 = J + B_1 + B_2 + K$$
$$E_2 = \tfrac{1}{2}(B_1 + B_2 + 2K) + \tfrac{1}{2}\sqrt{(B_1-B_2)^2 + 4J^2}$$
$$E_3 = \tfrac{1}{2}(B_1 + B_2 + 2K) - \tfrac{1}{2}\sqrt{(B_1-B_2)^2 + 4J^2}$$
$$E_4 = E_4(B_1, B_2, J, K)$$
$$E_5 = E_5(B_1, B_2, J, K) \qquad (29)$$
$$E_6 = E_6(B_1, B_2, J, K)$$
$$E_7 = \tfrac{-1}{2}(B_1 + B_2 - 2K) + \tfrac{1}{2}\sqrt{(B_1-B_2)^2 + 4J^2}$$
$$E_8 = \tfrac{-1}{2}(B_1 + B_2 - 2K) - \tfrac{1}{2}\sqrt{(B_1-B_2)^2 + 4J^2}$$
$$E_9 = J - B_1 - B_2 + K$$

where, in general, the values of $E_4$, $E_5$ and $E_6$ should be computed numerically.

The case ($K=0$, $J=1$ and $B_1 \neq B_2$) has been studied in Ref. [31] and the case ($B_1 = B_2$ and $K \neq 0$) has been studied in Refs. [32, 33]. In this paper we will consider the case ($J = -1$, $-2 \leq K \leq 0$ and $B_1 \neq B_2$). With these choices $|\Phi_2\rangle, ..., |\Phi_8\rangle$ are all entangled. In the two following sections, we will clarify why we consider the nonzero nonlinear coupling and the nonuniform magnetic field simultaneously.

## 6. Negativity of the System

In this section, we consider the negativity as a measure of entanglement and study its dependence on the system's parameters.

In Fig. 1, the negativity is plotted as a function of $B_1$ and $B_2$ for $K = -1.7$ and three typical temperatures. As it can be seen from Fig. 1(a), the negativity is zero in the most of the ($B_1 B_2 > 0$) region and is nonzero elsewhere. The reason is that for low temperatures the system is, almost, in its ground state and in the region where the negativity is zero, either $|\Phi_1\rangle$ or $|\Phi_9\rangle$ are ground states which are separable, while in the region with nonzero negativity, $|\Phi_4\rangle$ is the ground state which is entangled.

Figure 1 shows that the negativity in (almost all over) the $(B_1 B_2 > 0)$ region is also zero for higher temperatures despite the fact that in this case $|\Phi_1\rangle$ or $|\Phi_9\rangle$ will be mixed with entangled eigenstates of the Hamiltonian; but in the case where $|\Phi_4\rangle$ is the ground state, the negativity can remain nonzero for higher temperatures too. For other $K$ values, the situation is almost the same as above. This fact that the entanglement can be found usually in the $(B_1 B_2 < 0)$ region and not in the $(B_1 B_2 > 0)$ region is one of our reasons for considering the nonuniform magnetic field.

As we can see in Fig. 1, the negativity is symmetric with respect to $B_1 = \pm B_2$ lines. In addition, the peaks of the negativity are located on $B_1 = -B_2$ line. So in Fig. 2, we only represent the negativity on this line for $T = 1$ and three typical values of $K$. We observe that increasing $|K|$ can enhance the entanglement. This is one of our reasons for considering the nonzero $K$.

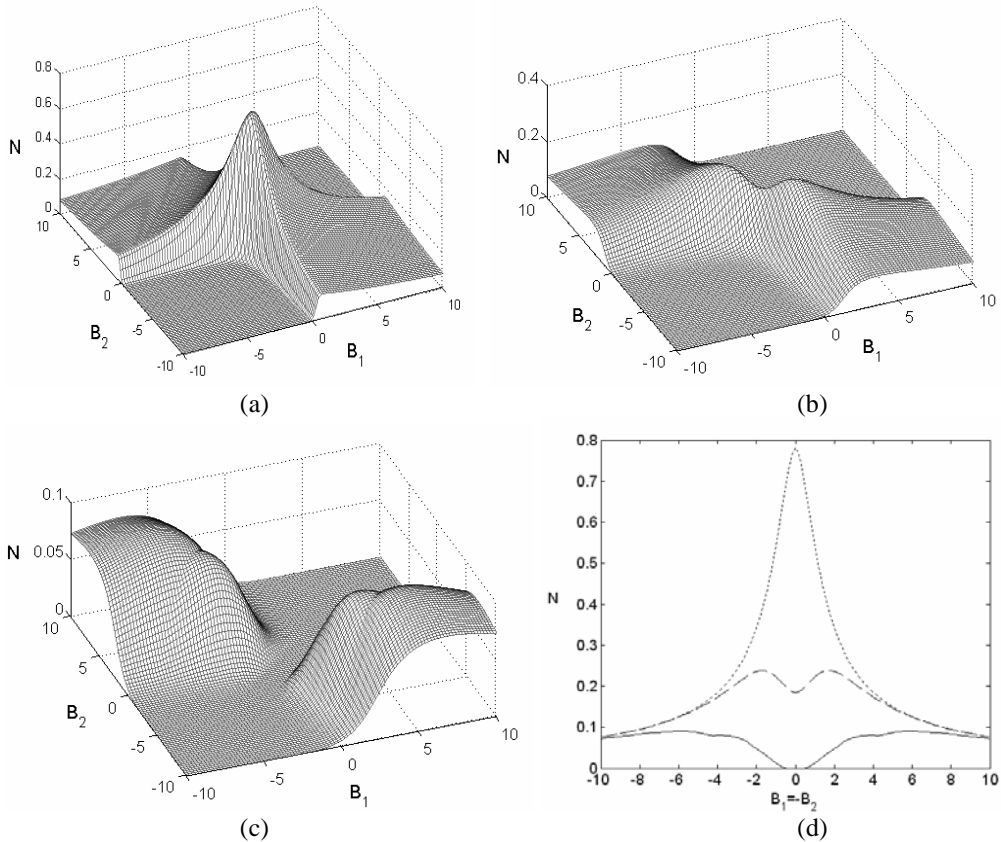

Fig. 1. Negativity versus $B_1$ and $B_2$ for $K = -1.7$ and three typical temperatures: (a) $T = 0.2$, b) $T = 1$, c) $T = 1.9$. d) Negativity on $B_1 = -B_2$ line for the three above cases: $T = 0.2$ (dotted line), $T = 1$ (dashed line) and $T = 1.9$ (solid line).

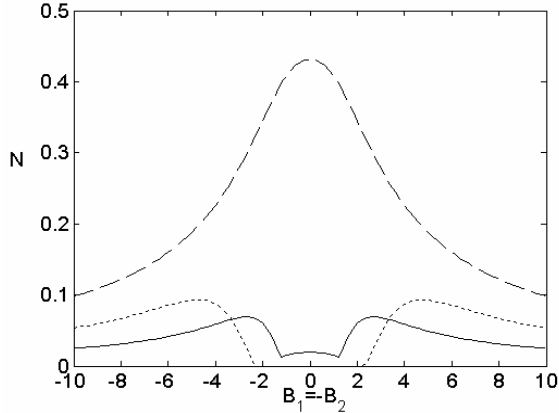

Fig. 2. Negativity on $B_1 = -B_2$ line for $T = 1$ and $K = -0.01$ (dotted line), $K = -1$ (solid line) and $K = -2$ (dashed line).

## 7. Increasing $|K|$ or $|B_1 - B_2|$ Can Increase $T_s$

Suppose that the ground state of the system is entangled and nondegenerate. This is the case for Figs. 3 and 4, since the ground state in these cases is $|\Phi_4\rangle$. By increasing the temperature, because of the mixing between the ground state and other eigenstates of the Hamiltonian, $\rho$ becomes separable. So it seems that any thing which prevents the mixing of the entangled ground state with other states may increase $T_s$. Any effect which increases the separation between the energy of the entangled ground state and other energy levels is one such effect. Increasing $|K|$ or $|B_1 - B_2|$ induces this effect in our case.

When $|\Phi_4\rangle$ is the ground state, our numerical calculations show that by increasing $|K|$, the separation between the ground energy level $E_4$ and other energy levels increases, i.e. for fixed $B_1$ and $B_2$, $E_i(|K|=|K_0|+\Delta) - E_4(|K|=|K_0|+\Delta) > E_i(|K|=|K_0|) - E_4(|K|=|K_0|)$ where $\Delta > 0$ and $i = 1,2,3,5,\ldots,9$ (so $\lambda_4(|K|=|K_0|+\Delta,T) > \lambda_4(|K|=|K_0|,T)$ for $0 < T < \infty$). In addition, by increasing $|B_1 - B_2|$, in most cases, the distance between the ground energy level $E_4$ and other energy levels either becoms larger or does not change. Since increasing $|K|$ or $|B_1 - B_2|$ can increase the gap between the ground energy level $E_4$ and other energy levels, we expect that increasing $|K|$ or $|B_1 - B_2|$ increases $T_s$. Figures 3 and 4 show that, with a good approximation, this guess is correct (see also [Fig. 1(d)]).

This line of reasoning for increasing $T_s$ is not accurate in general. However, it helps us to guess the behavior of $T_s$. As we have seen, it works well in our 2-qutrit system. We have also examined this approach for a few other simple systems and have obtained convincing results. For example, consider a 2-qubit anisotropic Heisenberg chain in the presence of uniform external magnetic field. The Hamiltonian and its eigenvalues are [34]:

$$H = J(S_1^+ S_2^- + S_1^- S_2^+) + J\gamma(S_1^+ S_2^+ + S_1^- S_2^-) + B(S_{1z} + S_{2z}) \tag{30-a}$$

$$E_1 = J, \ E_2 = -J, \ E_3 = \sqrt{B^2 + (J\gamma)^2}, \ E_4 = -\sqrt{B^2 + (J\gamma)^2} \tag{30-b}$$

where $S^\pm \equiv S_x \pm i S_y$, $S_{i\alpha} \equiv \frac{1}{2}\sigma_{i\alpha}$ ($\alpha = x, y, z$) in which $\sigma_{i\alpha}$ are the Pauli matrices at site i, $J \equiv (J_x + J_y)/2$, $B$ is the uniform external magnetic field along the z-axis and $\gamma \equiv (J_x - J_y)/(J_x + J_y)$. When $\gamma \neq 0$, all eigenvectors of $H$ are entangled [34]. We consider the case $\gamma = 0.8$, $J = 1$ and $B \geq 0$. For $0 \leq B < 0.6$, the ground state is $|\Phi_2\rangle = \frac{1}{\sqrt{2}}(|01\rangle - |10\rangle)$, which is the maximally entangled state, and $|\Phi_4\rangle$ is the first excited one (we have assumed that $H|\Phi_i\rangle = E_i|\Phi_i\rangle$). But for $B > 0.6$, $|\Phi_4\rangle$ is the ground state and $|\Phi_2\rangle$ is the first excited one and by increasing $B$, the difference between the ground energy level $E_4$ and the other energy levels increases. So we expect that for $B > 0.6$, increasing $B$ increases $T_s$. As one can see from Fig. 4 of Ref. [31], it is so for approximately $B \geq 2$. So our simple reasoning works almost satisfactory in this example. For $0.6 < B < 2$, the ground energy level $E_4$ is rather close to the first excited one. In addition, $|\Phi_2\rangle$ is more entangled than $|\Phi_4\rangle$. It seems that in such a situation our simple reasoning does not work, But when $E_4$ becomes sufficiently far from $E_2$, i.e. $B > 2$, it works well.

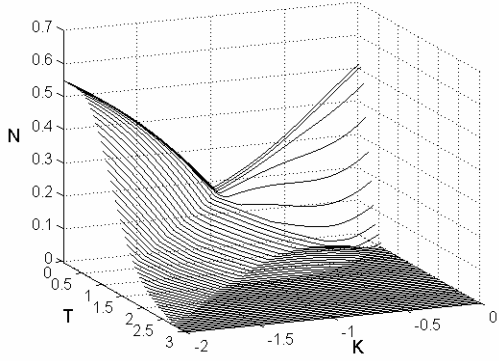 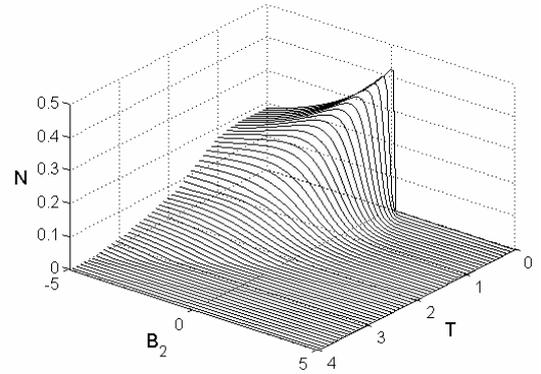

Fig. 3. Negativity versus $K$ and $T$ ($T$ begins at $T = 0.01$) for $B_1 = -B_2 = 1.3$.

Fig. 4. Negativity versus $B_2$ and $T$ ($T$ begins at $T = 0.02$) for $K = -1.7$ and $B_1 = 3$.

Note that as Figs. 3 and 4 show, increasing $|K|$ or $|B_1 - B_2|$, in fact, reduces the damping rate of the negativity by increasing the temperature. In addition, strictly speaking, since the negativity can not detect all entangled states in a 2-qutrit system, we should say that increasing $|K|$ or $|B_1 - B_2|$ can increase a *lower bound* of $T_s$. Do we get a similar result if we use another entanglement measure instead of the negativity? In the next section, we will see that the answer is yes, at least, if we use a lower bound of I-concurrence instead.

We also note that in Fig. 3, increasing $|K|$, for a fixed $T$, first decreases and then increases the negativity of $\rho(T)$. Specially for $T = 0$ case, increasing $|K|$ decreases the negativity up to $|K| = 1$; but for $|K| > 1$, increasing $|K|$ increases the negativity of the ground state. So for $|K| = 2$, the negativity of the ground state becomes even larger than its value for $|K| = 0$. Therefore, since increasing $|K|$ also increases $T_s$, Fig. 3 shows that by increasing $|K|$ we can achieve both larger negativity and higher $T_s$.

In Fig. 4, there is a jump in the amount of the negativity at ( $B_2 \approx 0.148$ and $T = 0$ ). This is because that $|\Phi_4\rangle$ is the ground state for $B_2 < 0.148$ but $|\Phi_9\rangle$ is the ground state for $B_2 > 0.148$. In contrast to Fig. 3, for $B_2 < 0.148$, increasing $|B_1 - B_2|$ only decreases the negativity of the ground state.

At the end of this section we summarize our reasons for considering nonzero $K$ and nonuniform magnetic field simultaneously:
1. Entanglement usually occurs in the $(B_1 B_2 < 0)$ region.
2. Increasing $|K|$ can enhance the entanglement.
3. Increasing $|K|$ or $|B_1 - B_2|$ can increase $T_s$.

## 8. Comparison of Two Lower Bounds of I-Concurrence

By choosing different $\{Z_\alpha\}$ in Eq. (9), we can obtain different lower bounds for $C(\rho)$. One simple way is to choose $Z_\kappa = 1$ ($\kappa = 1, \ldots, n$) and $Z_\alpha = 0$ for $\alpha \neq \kappa$. In this way we get "$n$" different lower bounds for $C(\rho)$. We call the best of these lower bounds the *algebraic lower bound* (ALB) of $C(\rho)$ [15]. Chen *et al.* [16] have introduced two lower bounds for $C(\rho)$, the one quoted in Eq. (7) and another one based on the realignment criterion. Since in our case the lower bound in Eq. (7) is often a better bound, we will only compare ALB with the bound in Eq. (7). Also note that, since Chen *et al.*'s lower bound in Eq. (7) is proportional to the negativity, comparing it with the ALB is in fact, the comparison between the negativity and the ALB.

In most of the cases which we have considered numerically, ALB gives a better lower bound for $C(\rho)$ than Eq. (7). Fig. 5(a) shows an example of this case. From Eq. (6), it is obvious that $UB(\rho) \equiv \sum_j \lambda_j C(|\Phi_j\rangle)$ is an upper bound of $C(\rho)$. The dotted line in Fig. 5(a) shows this upper bound. So for the case which is plotted in Fig. 5(a), the ALB can detect all entangled states. However, we have encountered in our calculations some cases where Eq. (7) gives a better lower bound for $C(\rho)$ than ALB [Fig. 5(b)].

But an important point, which is seen from Figs. 5(a) and 5(b), is that the behavior of these two lower bounds as a function of $B_1$ are almost similar. It is almost the case for other situations too. In other words, in our system, ALB and negativity usually behave similarly as a function of system's parameters. In particular, using ALB instead of negativity in Figs. 3 and 4 gives us similar results [see Figs 5(c) and 5(d)].

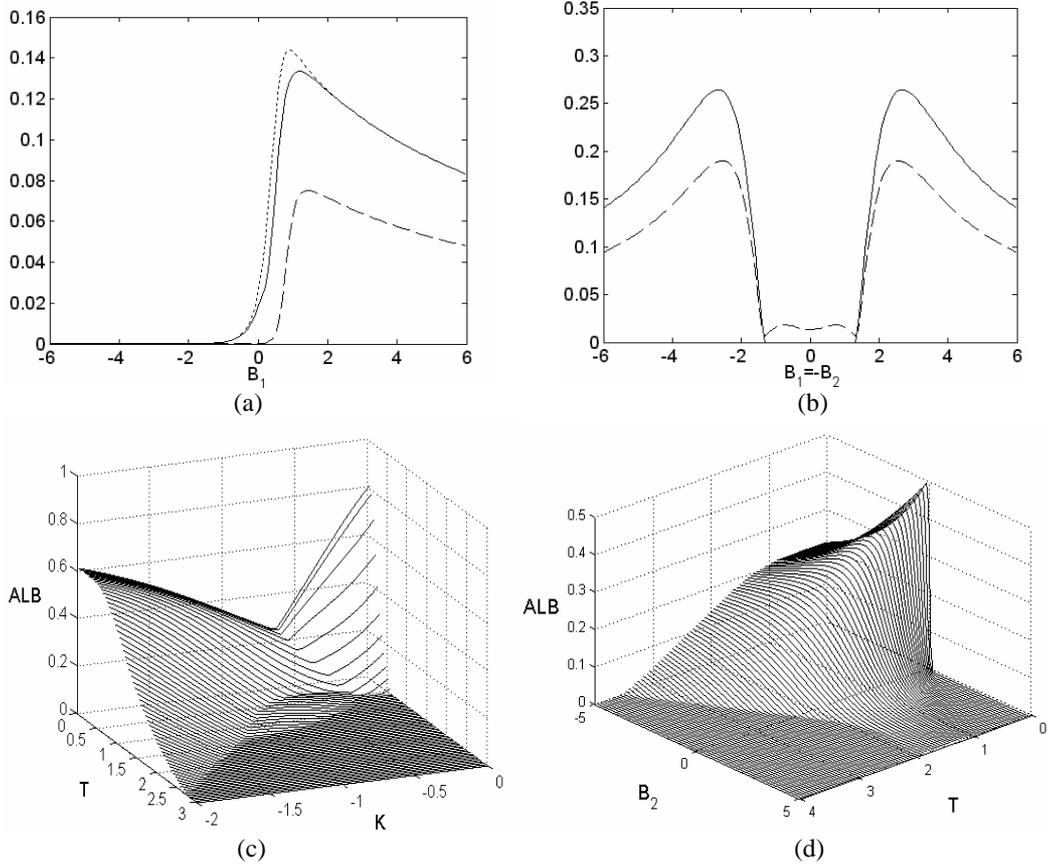

Fig. 5. (a) $UB(\rho)$ (dotted line) and two lower bounds of $C(\rho)$ (ALB, solid line; Eq. (7), dashed line) for $K = -1, T = 0.3$ and $B_2 = -6$. (b) Two lower bounds of $C(\rho)$ (ALB, solid line; Eq.(7), dashed line) for $K = -0.2$, $T = 0.5$ on $B_1 = -B_2$ line. (c) ALB versus $K$ and $T$ for the same conditions as Fig. 3 ($T$ begins at $T = 0.05$). (d) ALB versus $B_2$ and $T$ for the same conditions as Fig. 4 ($T$ begins at $T = 0.05$).

## 9. Dense Coding Capacity of the System

In this section we study the dense coding capacity of our 2-qutrit system. The case ($J=0$, $K \neq 0$ and $B_1 = B_2$) has been studied in Ref. [35]. As stated before in Sec. 5, in this paper we consider the case ($J = -1$, $-2 \leq K \leq 0$ and $B_1 \neq B_2$).

In Sec. 3, we have seen that a bipartite $\rho$ is useful for dense coding from 1 to 2 if $S(\rho_2) - S(\rho) > 0$, where $\rho_2 \equiv Tr_1(\rho)$ (in this section we denote subsystems by 1 and 2 instead of A and B). So one may define the *usefulness* of $\rho$ for dense coding from 1 to 2 and vice versa as

$$U_{DC:1;2}(\rho) = \max(S(\rho_2) - S(\rho), 0) \tag{31-a}$$

and

$$U_{DC:2;1}(\rho) = \max(S(\rho_1) - S(\rho), 0) \tag{31-b}$$

respectively.

Because of the term $B_1 S_{1z} + B_2 S_{2z}$ in Eq. (27), for our 2-qutrit system, the spectrum of $\rho_1$, in general, is not the same as that of $\rho_2$. So, in general, we have $S(\rho_1) \neq S(\rho_2) \Rightarrow U_{DC:2;1}(\rho) \neq U_{DC:1;2}(\rho)$.

$U_{DC:1;2}(\rho)$ for the same conditions as Fig. 3 is plotted in Fig. 6(a). Interestingly, the dependence of $U_{DC:1;2}(\rho)$ on $K$ and $T$ is similar to that of negativity in Fig. 3. In addition, for the case plotted in Fig. 6(a), we have $U_{DC:2;1}(\rho) = U_{DC:1;2}(\rho)$. But for the same conditions as Fig. 4, in general, we have $U_{DC:2;1}(\rho) \neq U_{DC:1;2}(\rho)$ as can be seen from Figs. 6(b) and 6(c). Figures 4, 6(b) and 6(c) show that though the behavior of $U_{DC:1;2}(\rho)$ as a function of $B_2$ and $T$, is not similar to that of $N(\rho)$, this is so for $U_{DC:2;1}(\rho)$. In fact, for the all cases which we have considered numerically, the behavior of $\max[U_{DC:1;2}(\rho), U_{DC:2;1}(\rho)]$, which is a lower bound of $E_R(\rho)$, has been found to be similar to that of $N(\rho)$.

It seems interesting that there are states which in spite of being distillable, i.e. $\max[U_{DC:1;2}(\rho), U_{DC:2;1}(\rho)] > 0$, they are useless for dense coding from one of the two parts to the other, e.g. from 1 to 2 and for $(B_2 = -10, T = 1.5)$, as Figs. 6(b) and 6(c) show. Also, Fig. 6(d) shows that there are states though are NPT, but are useless for dense coding (for this case we have $U_{DC:2;1}(\rho) = U_{DC:1;2}(\rho)$). In Fig. 6(d), we have plotted $C_{DC}(\rho)$ instead of $U_{DC:1;2}(\rho)$. Since for pure product states we have $C_{DC}(\rho) = \log_2 d = \log_2 3$, Fig. 6(d) shows that there are NPT entangled states which are even weaker than pure product states for encoding information by manipulating only one part of the total system.

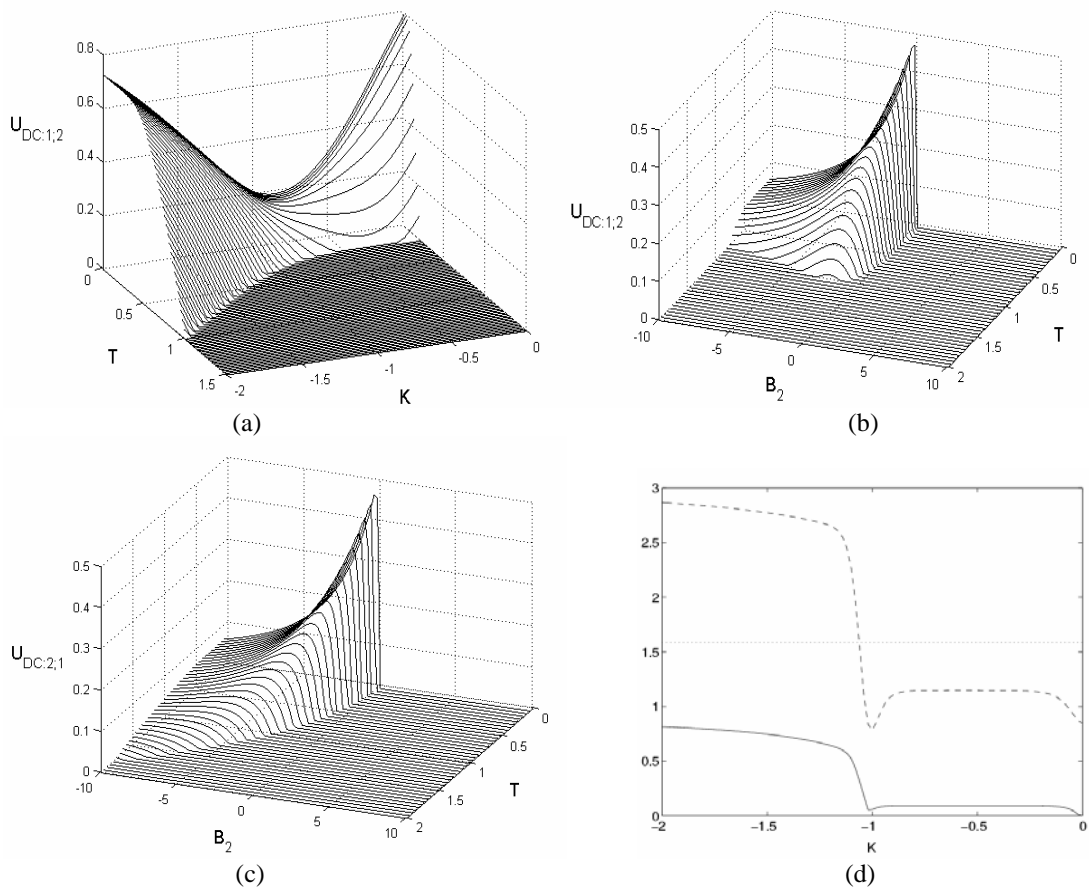

Fig. 6. (a) $U_{DC:1;2}$ versus $K$ and $T$ for the same conditions as Fig. 3 ($T$ begins at $T = 0.02$). (b) $U_{DC:1;2}$ versus $B_2$ and $T$ for the same conditions as Fig. 4 ($T$ begins at $T = 0.05$). (c) $U_{DC:2;1}$ versus $B_2$ and $T$ for the same conditions as Fig. 4 ($T$ begins at $T = 0.05$). (d) $C_{DC}$ (dashed line) and the negativity (solid line) versus $K$ for $B_1 = B_2 = 0$ and $T = 0.05$. The dotted line corresponds to $y = \log_2 3$.

## 10. Conclusion

By using the fact that the maximally mixed state is surrounded by separable states, it can be shown that thermal entanglement of an arbitrary (finite dimensional) $m$-partite system vanishes at a finite temperature $T_s$. In our 2-qutrit system, entanglement usually occurs in the $(B_1 B_2 < 0)$ region and increasing $|K|$ or $|B_1 - B_2|$ can increase $T_s$.

The interesting feature of our system is that, for a fixed *T*, increasing $|K|$ can also enhance the entanglement of $\rho(T)$; in particular it can enhance the entanglement of the ground state.

We have also compared the negativity with the algebraic lower bound (ALB) of $C(\rho)$. We have observed that they behave almost similarly as a function of system's parameters. In particular, using ALB instead of negativity confirms that increasing $|K|$ or $|B_1 - B_2|$ can increase $T_s$.

Studying the dense coding capacity of the system, we have seen that the effect of increasing $|K|$ or $|B_1 - B_2|$ on the $\max[U_{DC:1;2}, U_{DC:2;1}]$ is similar to its effect on the negativity of the system.

As we have seen, manipulating the system's parameters (including temperature), gives us the ability to tune the amount of entanglement of the system. This ability may lead to interesting properties for systems with more than two parts [36]. In addition, studying the thermal entanglement of *n*-qutrit spin chain (with $n > 2$), may help us to know whether a monogamy inequality, specially in terms of the negativity, is possible for qudits, with $d > 2$, or not[a].

## Acknowledgments

We would like to thank the anonymous reviewer for his/her helpful comments and suggestions. We also thank J. Jannati and R. A. Zeinali for their helps in MATLAB.

---

[a] Entanglement in *n*-qubit systems (with $n > 2$) is monogamous; i.e. there is a trade-off between the amount of entanglement between qubits $A_1$ and $A_2$ with the amount of entanglement between $A_1$ and other $A_i$, ($i \neq 2$) [2, 37, 38]. The monogamy property of the entanglement between qubits, leads to inequalities in terms of the tangle [37, 38] and also the negativity [39]. For higher-dimensional systems, i.e. qudits with $d > 2$, the monogamy inequality in terms of tangle is no longer valid [40], and it is not known whether the monogamy inequality in terms of negativity is valid or not. It is however worthnoting that for qudits, there is a monogamy inequality for, at least, pure states [41].